\documentclass{PoS}

\newcommand{\pbp}{\bar{\psi}\psi}
\newcommand{\Tspace}{\rule{0pt}{3ex}}
\newcommand{\Bspace}{\rule[-1.2ex]{0pt}{0pt}}

\title{Dirac Eigenvalue Spectrum at Finite 
Temperature Using Domain Wall Fermions}

\ShortTitle{Dirac Eigenvalue Spectrum at Finite Temperature Using DWF}

\author{\speaker{Zhongjie Lin} (HotQCD Collaboration)%
\\Department of Physics, Columbia University, New York, NY 10027, USA\\
E-mail: \email{jasper@phys.columbia.edu}}

\abstract{%
We present a study of the Dirac eigenvalue spectrum near the region of 
the QCD phase transition. This study makes use of a sequence of ensembles 
with temperatures from 150 MeV to 200 MeV generated with $2+1$ flavors of 
dynamical domain wall fermions (DWF) and the dislocation suppressing 
determinant ratio (DSDR) action on a $16^3\times8$ lattice with an extent 
of $32$ or $48$ in the fifth dimension. All the simulations lie on a 
line of constant physics with 200 MeV pions. The DWF Dirac operator is 
normalized using the methods of Giusti and Luscher combined with those of 
Rome-Southampton collaboration, allowing a direct evaluation of the 
Banks-Casher relation.  The relation between the resulting
temperature-dependent Dirac eigenvalue spectrum and the possible 
restoration of $U(1)_A$ symmetry with increasing temperature is discussed.
}

\FullConference{XXIX International Symposium on Lattice Field Theory \\
July 10-16, 2011\\
Squaw Valley, Lake Tahoe, California}

\begin{document}

\section{Introduction}
The chiral phase transition is one of the most fundamental features of QCD.
Lattice field theory has been applied successfully to the study of this 
interesting phenomena and the associated symmetries. While traditional
lattice techniques measure the chiral observables in a straightforward 
manner, examining the low-lying part of the eigenvalue spectrum of the 
Dirac operator can provide unique insights into various aspects of the 
symmetry breaking or restoration that accompany the phase transition. 

For instance, the chiral condensate, the chiral order parameter for the 
transition, can be expressed in terms of the eigenmode density via the 
following relation:
\begin{equation}
  -\langle\bar{\psi}_q\psi_q\rangle=\int\mathrm{d}\lambda\,\rho(\lambda)
  \frac{2m_q}{m_q^2+\lambda^2}\,,\qquad q=l,s,
  \label{eqn:pbp}
\end{equation}
where $\rho(\lambda)$ is the spectral density of the Dirac operator and
$m_q$ is the quark mass.  When the chiral and infinite-volume limits are 
taken, one will obtain the well-known Banks-Casher 
relation~\cite{Banks:1979yr}, 
\begin{equation}
  \lim_{m_l\to0}\lim_{V\to\infty}
  -\langle\bar{\psi}_l\psi_l\rangle
  =\pi\lim_{\lambda\to0}\lim_{m_l\to0}\lim_{V\to\infty}\rho(\lambda).
  \label{eqn:bcr}
\end{equation}

In lattice calculation one may also examine the subtracted 
chiral condensate defined as,
\begin{equation}
  \Delta_{l,s}= 
  \langle\bar\psi_l\psi_l\rangle - 
  \frac{m_l}{m_s}\langle\bar\psi_s\psi_s\rangle \; .
  \label{eqn:sub}
\end{equation}
The subraction removes the ultraviolet divergent piece of the chiral 
condensate which is linear in quark mass.

The Dirac eigenvalue spectrum can be utilized to study the anomalous 
$U(1)_A$ symmetry as well. A similar order parameter $\Delta_{\pi-\delta}$,
which is the difference between the pseudoscalar and scalar 
susceptibilities, can also be related to the eigenvalue spectrum,
\begin{equation}
  \Delta_{\pi-\delta} \equiv \frac{\chi_{\pi} - \chi_{\delta}}{T^2}
  = \frac{1}{T^2}\int\mathrm{d}\lambda\ \rho(\lambda)
  \frac{4m_l^2}{\left(m_l^2+\lambda^2\right)^2}\; .
  \label{eqn:u1a}
\end{equation}
The expression above suggests that if there is a finite region above zero 
where the eigenvalue density vanishes (a gap), the $U(1)_A$ symmetry 
might be effectively restored.

With chiral symmetry under good control, domain wall fermions 
(DWF)~\cite{Kaplan:1992bt, Furman:1994ky} are an optimum tool for the 
exploration of the phase transition region. The residual chiral symmetry 
breaking (present in the DWF formalism for finite fifth-dimensional extent 
$L_s$) is reflected in an additive correction to the bare quark mass 
($m_\mathrm{res}$), which can be further suppressed by the adoption of the 
dislocation suppression determinant ratio (DSDR) 
action~\cite{Vranas:1999rz,Fukaya:2006vs,Renfrew:2009wu}. 
Despite some unphysical, massive degrees of freedom from the extra fifth 
dimension, the low modes of the DWF Dirac operator should resemble an
ordinary four-dimensional discretized version of QCD. Moreover, within the 
DWF formalism, the $U(1)_A$ symmetry is only broken by axial anomaly, 
rather than spoiled by lattice artifacts as with staggered fermions.

\section{Implementation Details}
We have collected eight ensembles near the phase transition region with 
$2+1$ flavors of fermions. All the simulations have a $16^3\times8$ 
space-time volume and a fifth dimension of $L_S=32$ or $48$ and they
all lie on a line of constant physics with $m_\pi\approx 200 \mathrm{MeV}$ 
and kaon mass almost physical~\cite{Cheng:2011lat}.  Table~\ref{tab:sum} 
gives the basic parameters of these finite-temperature ensembles. The 
$N_{\rm cfg}$ column lists the number of configurations for which the 
eigenvalues are calculated. Figure~\ref{fig:chi} shows the disconnected 
susceptibilities of various temperatures and indicates a critical
temperature around 160 MeV.
\begin{table}[h]\footnotesize
  \centering
  \begin{tabular}{ccc|ccc|ccc|c|c} \hline
    $T\,(\mathrm{MeV})$&$\beta$&$L_s$&$m_{\mathrm{res}}a$
    &$m_la$&$m_sa$&
    $\Tspace \Bspace \frac{\left<\pbp\right>_l}{T^3}$&
    $\frac{\Delta\pbp}{T^3}$&$\frac{\chi_{l,{\rm disc}}}{T^2}$ &$N_\mathrm{cfg}$&
    $Z_{{\rm tw} \to m_f}^{(\pi)}$\\ \hline
    140&1.633&48&0.00612&-0.00136&0.0519&6.26(12)& 7.74(12)&36(3)&  -&       -\\ 
    150&1.671&48&0.00296& 0.00173&0.0500&6.32(29)& 6.10(29)&41(2)&340&1.980(7)\\ 
    150&1.671&32&0.00648&-0.00189&0.0464&8.39(10)& 7.06(10)&44(3)&340&1.905(6)\\ 
    160&1.707&32&0.00377&0.000551&0.0449&5.25(17)& 4.83(17)&43(4)&408&1.725(8)\\
    170&1.740&32&0.00209&0.00175 &0.0427&4.03(18)& 2.78(18)&35(5)&239&1.631(11)\\
    180&1.771&32&0.00132&0.00232 &0.0403&3.16(15)& 1.56(15)&25(4)&246&1.476(4)\\
    190&1.801&32&0.00076&0.00258 &0.0379&2.44(9) & 0.71(9) &11(4)&374&1.439(3)\\
    200&1.829&32&0.00046&0.00265 &0.0357&2.19(8) & 0.47(8) &10(3)&710&1.365(3)\\
    \hline
    0  &1.750&32&0.00188&0.00300 &0.0370&       -&        -&    -&252&1.5685(5)\\
    \hline
  \end{tabular}
  \caption{Summary of ensembles and the renormalization factors 
  for the eigenvalue density.}
  \label{tab:sum}
\end{table}

\begin{figure}[htb]
  \begin{center}
	\begin{minipage}[t]{0.5\linewidth}
		\centering
        \resizebox{\linewidth}{!}{\input{./figs/Disc_Susc.tex}}
        \caption{Disconnected susceptibilites.}
        \label{fig:chi}
	\end{minipage}% 
    \begin{minipage}[t]{0.5\linewidth}
		\centering
        \resizebox{\linewidth}{!}{\input{./figs/ratio_b175.tex}}
        \caption{Renormalization factors for the $\beta=1.750,\; 16^3\times16$ ensemble.}
        \label{fig:npr}
	\end{minipage}
  \end{center} 
\end{figure}

For comparison, results from a zero-temperature ensemble with a volume of 
$16^3\times16$ are presented as well.  In order to keep $m_\pi=200$ MeV as 
the temperature and $\beta$ decreases and $m_{\rm res}$ grows we must 
either increase $L_s$ above 32 or use a negative input quark mass.  As the 
second and third lines of Table~\ref{tab:sum} show, this first use of a 
negative DWF input quark mass was successful, resulting in no exceptional 
configurations and giving a consistent result for $\chi_{l,{\rm disc}}$.

We used the Kalkreuter-Simma \cite{Kalkreuter:1995mm} method to calculate 
the lowest $N_\mathrm{eig}=100$ eigenvalues of the Hermitian version of the
Dirac operator $D_H\equiv R^5\gamma^5D_\mathrm{DWF}$, where $R^5$ is the 
reflection operator in the fifth dimension.  Note the eigenvalues measured 
here include the mass term and are denoted by $\Lambda$, to be 
distinguished from those in the equations~\ref{eqn:pbp}, \ref{eqn:bcr} 
and \ref{eqn:u1a}.

\section{Eigenvalue spectrum and its renormalization}
Renormalization must be applied to the eigenvalue spectrum of the DWF 
operator $D_H$ before any sensible comparison can be made with either the 
input DWF quark mass $m_f$ or the eigenvalue densities from other 
four-dimensional fermion formalisms ({\it e.g.} Wilson fermions).
The method we have adopted is a generalization to DWF of that proposed by 
Giusti and Luscher~\cite{Giusti:2008vb}, which introduces into the 
Lagrangian a twisted mass term,
\begin{equation}
  {\cal L}_\mathrm{tm}(x) = \sum_{j=1}^k \overline{q}^j(x)
  \left(\gamma^\nu D_\nu + m + i\mu\gamma^5\tau^3\right)q^j(x) .
  \label{eqn:tml}
\end{equation}
Then Green's functions such as the six-point correlator of the charged 
pseudoscalar density operator
$P^\pm_{ll'} = \overline{q}\,^l(x)\gamma^5\tau^\pm q^{l'}(x)$
can be expressed in terms of the eigenvalue density,
\begin{eqnarray}
  \sigma_3(\mu)&=&
  -\sum_{x_i}\left\langle P^+_{1,2}(x_1) P^-_{2,3}(x_2)
  P^+_{3,4}(x_3) P^-_{4,5}(x_4)P^+_{5,6}(x_5) P^-_{6,1}(x_6)
  \right\rangle\\
  &=&
  \left\langle{\rm Tr}\left\{
  \frac{1}{\left((\gamma^5 D)^2 + \mu^2\right)^3}
  \right\}\right\rangle\\
  &=& 
  \int_{-\infty}^\infty \mathrm{d}\,\Lambda \rho(\Lambda)
  (\Lambda^2 + \mu^2)^{-3},
  \label{eqn:sigma_3}
\end{eqnarray}
in the notation of Giusti and Luscher~\cite{Giusti:2008vb}.  The charged 
pseudoscalar density is well defined in the continuum limit in a variety of
regularization schemes which are related by multiplicative renormalization 
factors.  Equation \ref{eqn:sigma_3} implies the same rules also will work 
for the eigenvalue density {\it e.g.}:
\begin{equation}
  P_{ll'}^{\prime i} = \frac{1}{Z_{m \to m'}} P_{ll'}^i\quad
  \Longrightarrow\quad
  \rho^\prime(\Lambda^\prime)=
  Z_{m\to m^\prime}^{-1}\rho\left(
  \frac{\Lambda^\prime}{Z_{m\to m^\prime}}\right)
  \label{eqn:mm}
\end{equation}
where $Z_{m\to m^\prime}$ is the renormalization factor for the mass term 
treated symmetrically with $P^\pm_{ll'}$.

With such inspiration, we can invent a five-dimensional analogue of the 
twisted-mass term 
$P_{ll'}^{{\rm DWF},i}(x) = \sum_{s=0}^{L_s-1}
\overline{\Psi}_l(x,s)\gamma^5\tau^i\Psi_{l'}(x,L_s-1-s)$
and relate it to the usual pseudoscalar density,
\begin{equation}
  \overline{\psi}(x) \gamma^5 \psi(x)
  \approx \frac{1}{Z_{{\rm tw} \to m_f}}
  \sum_{s=0}^{L_s-1}\overline{\Psi}(x,s)\gamma^5\Psi(x,L_s-1-s),
  \label{eqn:ps_equiv}
\end{equation}
where $\psi(x)$ is the four-dimensional operator while $\Psi(x, s)$ is the 
five-dimensional field.
\footnote{The explicit sum over the fifth ($s$) dimension is suppressed
later on for simplicity if no confusion is caused.}
This renormalization factor $Z_{{\rm tw} \to m_f}$ connects the 
five-dimensional eigenvalue density to a more conventional density 
normalized in a fashion consistent with the usual bare quark mass $m_f$:
\begin{equation}
  \rho^{m_f}(\Lambda^{m_f})=
  Z_{{\rm tw} \to m_f}^{-1}\rho(\Lambda^{(5d)}),\qquad
  \Lambda^{m_f}=Z_{{\rm tw} \to m_f}\Lambda^{(5d)}\,.
  \label{eqn:4d5d}
\end{equation}

Because of the equivalence of the two operators at long distance
expressed by equation~\ref{eqn:ps_equiv}, the renormalization 
factor can be obtained from the ratio of the two Green's functions:
\begin{equation}
  Z_{{\rm tw} \to m_f}^{(\pi)}
  = \frac{\left\langle \sum_{\vec x} 
  \overline{\Psi}(\vec x,t)R_5\gamma^5\tau^i\Psi(\vec x,t )O_\pi^i(0)\right\rangle}
  {\left\langle \sum_{\vec x}
  \overline{\psi}(\vec x,t)\gamma^5\tau^i\psi(\vec x,t)O_\pi^i(0)\right\rangle},
  \label{eqn:zpi}
\end{equation}
where $\tau_i$'s are the Pauli matrices in the flavor space.  Results from 
a Coulomb gauge fixed wall source are presented in Table~\ref{tab:sum}.

An alternative approach is to examine the off-shell, three-point Green's 
functions evaluated in Landau gauge.  This is very similar to the 
Rome-Southampton non-perturbative renormalization (NPR) technique 
(RI/MOM)~\cite{Martinelli:1994ty}.  The renormalization factor is extracted
from the ratio of amputated vertices for the five- and four-dimentional 
operators.
\begin{equation}
  Z_{{\rm tw} \to m_f}^{\rm(MOM)}(p_1,p_2)
  = \frac{ {\rm Tr}\left\langle \sum_{x_1,x_2} e^{i(p_2x_2-p_1x_1)} 
  \psi_l(x_2)\overline{\Psi}_l(0)R_5\gamma^5\Psi_{l'}(0)
  \overline{\psi}_{l'}(x_1)\right\rangle}
  { {\rm Tr}\left\langle \sum_{x_1,x_2} e^{i(p_2x_2-p_1x_1)}
  \psi_l(x_2)\overline{\psi}_l(0)\gamma^5\psi_{l'}(0)
  \overline{\psi}_{l'}(x_1)\right\rangle}.
  \label{eqn:zmom}
\end{equation}
To fully utilize the whole lattice, we use a series of fixed-momentum 
volume sources to calculate the propagators, which is defined as
\begin{equation}
  \eta(x\,;\,p)=e^{ip\cdot x}\,\mathbb{I}_{4\times4}\otimes\mathbb{I}_{3\times3}
  \; .
  \label{eqn:src}
\end{equation}
We perform our calculation using both non-exceptional kinematics, where
$p_1^2=p_2^2=(p_1-p_2)^2$, and exceptional kinematics, where $p_1=p_2$.
The results for the zero-temperature ensembles are presented in 
Table~\ref{tab:sum} and Figure~\ref{fig:npr}.

Both Equation~\ref{eqn:zpi} and \ref{eqn:zmom} should give consistent 
results independent of temporal separation $t$ and of $p_1$ and $p_2$ 
respectively.  Unfortunately, the NPR calculation is not feasible for the 
finite temperature ensembles due to large fluctuations. Therefore we only 
present the NPR results for the $16^3\times16$ emsemble in 
Figure~\ref{fig:npr}.  Figure~\ref{fig:npr} shows a discrepancy between the
two kinematics which is positively related to the physical momenta. This 
contradicts our expectation but can be plausibly explained by appreciable 
finite lattice spacing errors $(ap)^2$ at large momentum.  Because the 
quantity $Z^{(\pi)}_{{\rm tw}\to m_f}$ involves the smallest momenta, we 
use it to renormalize the spectrum.

Further renormalzation from the bare $m_f$ scheme to the conventional, 
continuum $\overline{\rm MS}$ scheme can then be easily performed since 
this final step has already been studied in detail~\cite{Aoki:2010dy}. 
Technical details and updated results will be available in our upcoming 
paper~\cite{Cheng:2011}.

\begin{figure}[h]
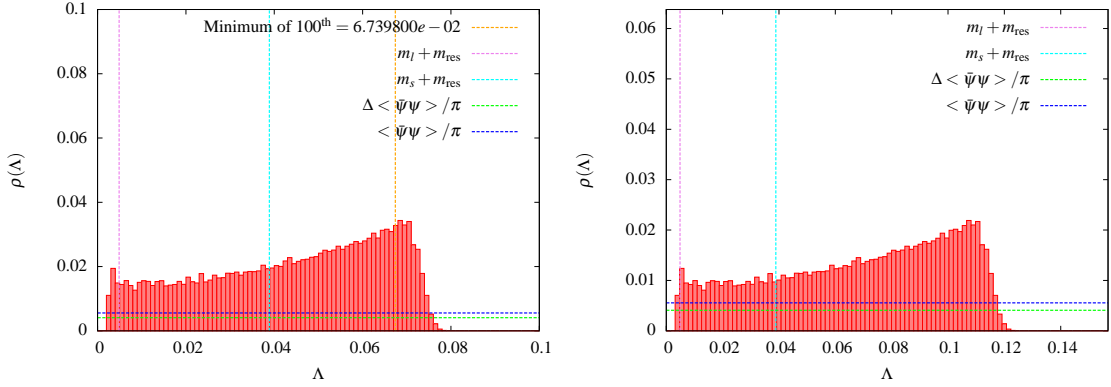

  \begin{center}
    \begin{minipage}[t]{0.5\linewidth}
      \centering
      \resizebox{\linewidth}{!}{\input{./figs/000MeV_ml003.tex}}
    \end{minipage}% 
    \begin{minipage}[t]{0.5\linewidth}
      \centering
      \resizebox{\linewidth}{!}{\input{./figs/000MeV_ml003_norm.tex}}
    \end{minipage}
  \end{center} 
  \label{fig:0mev}
  \caption{Dirac eigenvalue spectrum for the $\beta=1.750,\; 16^3\times16$,
  zero temperature ensemble.  The density in the left-hand panel has not 
  been renormalized while that on the right has been changed into the 
  normalization scheme of the usual input DWF mass $m_f$.  The left-most 
  vertical line locates the total bare quark mass, $m_f+m_{\rm res}$ which 
  matches well with the small peak seen in the renormalized, right-hand 
  spectrum.}
\end{figure}

Figure~\ref{fig:0mev} displays the effects of the renormalization, which
can be naively regarded as a rescaling of the axes.  The orange vertical 
line denotes the smallest of the hundredth eigenvalue and the spectrum 
below that is supposed to be complete. The other vertical lines indicate 
the bare masses of the light and strange quarks. The horizontal lines are 
the chiral condensates divided by $\pi$, which should agree with the 
eigenvalue density at $\lambda=0$ as predicted by Banks-Cahser 
relation~\ref{eqn:bcr}.  There are two significant features associated with
the renormalization. First, the light quark mass now matches the likely 
zero-mode peak.  Second, the Banks-Casher relation agrees better although 
it is still inaccurate at $30\%$ level. We attribute the discrepancy to  
finite-volume and finite-mass effects. Thus, no definitive conclusion can 
be drawn before studies on a larger lattice and the chiral extrapolation 
are performed.

\begin{figure}[htb]
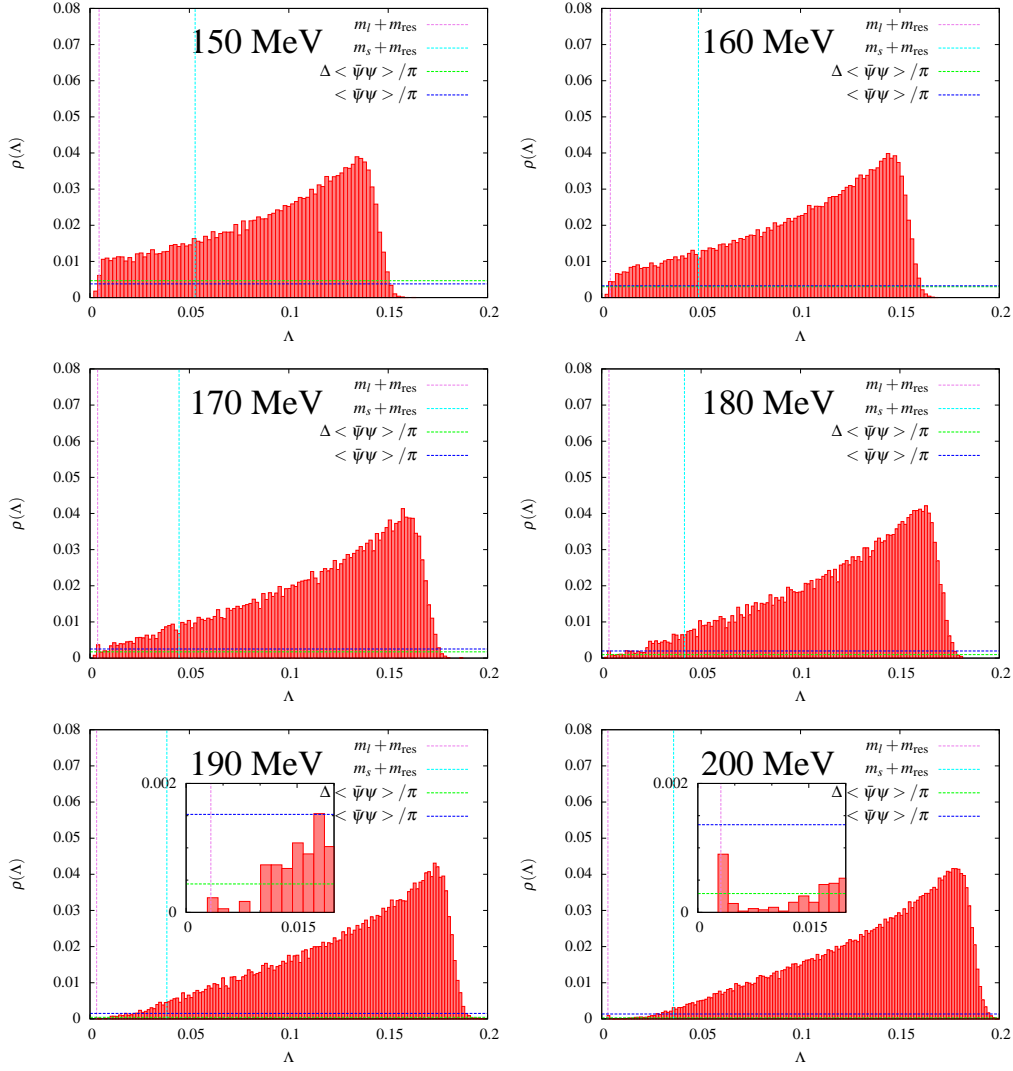

  \centering
  \begin{minipage}[t]{0.45\linewidth}
    \centering
    \resizebox{\linewidth}{!}{\input{./figs/150MeV_1_norm.tex}}
  \end{minipage}%
  \begin{minipage}[t]{0.45\linewidth}
    \centering
    \resizebox{\linewidth}{!}{\input{./figs/160MeV_norm.tex}}
  \end{minipage}
  \begin{minipage}[t]{0.45\linewidth}
    \centering
    \resizebox{\linewidth}{!}{\input{./figs/170MeV_norm.tex}}
  \end{minipage}% 
  \begin{minipage}[t]{0.45\linewidth}
    \centering
    \resizebox{\linewidth}{!}{\input{./figs/180MeV_norm.tex}}
  \end{minipage}
  \begin{minipage}[t]{0.45\linewidth}
    \centering
    \resizebox{\linewidth}{!}{\input{./figs/190MeV_norm.tex}}
  \end{minipage}%
  \begin{minipage}[t]{0.45\linewidth}
    \centering
    \resizebox{\linewidth}{!}{\input{./figs/200MeV_1_norm.tex}}
  \end{minipage}
  \caption{Dirac eigenvalue spectrum for the $T=150 - 200$ MeV ensembles.
  Here the temperature is lowest in the upper left and largest in the lower
  right.  The chiral symmetry breaking density of near zero eigenvalues
  disappears rapidly with increasing temperature and for the two highest
  temperature cases there appears to be a gap with very few eigenvalues
  just above zero.  The magnified inset in these two cases show some near
  zero eigenvalues and a suggestive zero mode peak located at 
  $\Lambda=m_f+m_{\rm res}$}
  \label{fig:eig}
\end{figure}

Figure~\ref{fig:eig} shows the renormalized eigenvalue spectra at various 
temperatures near the phase transition region. Although not in perfect
agreement with the Banks-Casher relation, at lower temperatures such as 150 
and 160 MeV the chiral condensates and the eigenvalue densities are 
different from zero, signaling spontaneous chiral symmetry breaking.  Above
170 MeV, these two quantities both start to vanish as expected for 
temperatures above the transition. However, it remains uncertain whether 
the slope of the eigenvalue density vanishes before 190 and 200 MeV, where 
a possible gap does begin to emerge, indicating an effective restoration of
$U(1)_A$ symmetry. A small peak at the lower end at these temperatures 
suggests that the major contribution to $U(1)_A$ symmetry breaking may come
from zero-modes, which are expected to go away as the volume increases. 
Therefore, we await a calculation on a larger lattice to confirm this 
conclusion.

\section{Conclusions}
With the chirally symmetric DWF framework, we are able to explore the
chiral and $U(1)_A$ symmetries near the phase transition region.
The successfully renormalized eigenvalue spectrum of the Dirac operator
as well as the correlator measurements~\cite{Hegde:2011lat} suggest an 
effective restoration of the $U(1)_A$ symmetry at temperatures higher 
than $T_c$.  We are looking forward to a similar simulation with larger 
volume to confirm our findings.

I very much appreciate the help and advice from members of HotQCD and my 
colleagues at Columbia University. This work was supported in part by U.S. 
DOE grant DE-FG02-92ER40699. The simulations were carried out on the BG/P 
machine at LLNL, the DOE- and RIKEN-funded QCDOC machines and NYBlue 
machine at BNL.

\end{document}